# Reliable and Cost-Efficient IoT Connectivity for Smart Agriculture: A Comparative Study of LPWAN, 5G, and Hybrid Connectivity Models

Mohamed Shabeer Mohamed Rafi[1], Mehran Behjati[1*] and Ahmad Sahban Rafsanjani[1]

[1] Department of Smart Computing and Cyber Resilience, Faculty of Engineering and Technology, Sunway University, Sunway 47500, Malaysia
`mehranb@sunway.edu.my`

**Abstract.** The integration of the Internet of Things (IoT) in smart agriculture has transformed farming practices by enabling real-time monitoring, data-driven decision-making, and automation. However, ensuring reliable connectivity in diverse agricultural environments remains a critical challenge. This paper analyzes the performance trade-offs between Low-Power Wide-Area Networks (LPWAN)—specifically LoRaWAN, NB-IoT, and Sigfox—and cellular networks (4G/5G) in agricultural applications. Beyond a comprehensive literature review (2020–2024), this study evaluates hybrid LPWAN–5G architectures that integrate the strengths of both network types to enhance cost-efficiency and connectivity reliability. Using real-world case studies, the findings demonstrate that hybrid LPWAN–5G models can reduce connectivity costs by up to 30% while significantly improving network reliability in remote agricultural settings. This work provides actionable recommendations for selecting optimal IoT connectivity solutions based on agricultural requirements and proposes future research directions to further optimize IoT infrastructure in smart farming.

## 1 Introduction

The global agricultural sector is undergoing a transformative shift driven by the adoption of smart agriculture technologies, which leverage the Internet of Things (IoT) to optimize resource use, enhance productivity, and promote sustainability [1]. With the world's population projected to reach 9.7 billion by 2050, the demand for food is expected to increase by 50%, necessitating innovative solutions to address challenges such as climate change, resource scarcity, and labor shortages [2]. Smart agriculture, enabled by IoT, offers a promising pathway to meet these challenges by providing real-time monitoring, data-driven decision-making, and automation across agricultural operations [3]. However, the success of smart agriculture hinges on reliable and cost-effective connectivity solutions that can operate seamlessly across diverse agricultural environments, from remote farms to controlled indoor settings. Connectivity is the backbone of IoT-enabled agriculture, enabling the collection, transmission, and analy-sis of data from sensors, drones, and other smart devices [4]. Without robust and adaptive network infrastructure, the full potential of IoT in agriculture cannot be fully realized, particularly in remote or challenging environments where connectivity is limited.

Two major categories of IoT connectivity have emerged in smart agriculture: Low-Power Wide-Area Networks (LPWAN) and cellular networks. LPWAN technologies such as Long-Range Wide Area Network (LoRaWAN), Narrowband IoT (NB-IoT), and Sigfox are optimized for long-range, low-bandwidth communication with minimal power consumption. In contrast, cellular networks (4G/LTE and 5G) provide higher data rates and low-latency connectivity but require more infrastructure investment and energy [5]. Each technology presents trade-offs in coverage, power consumption, data throughput, and cost, influencing their suitability for different agricultural applications. Recent research suggests that hybrid models combining LPWAN with 5G may offer an optimal balance between energy efficiency, data transmission reliability, and scalability [6]. However, the feasibility and effectiveness of these hybrid architectures require further investigation.

Despite extensive research on IoT connectivity in agriculture, selecting the most suitable network architecture remains a key challenge, particularly when balancing cost, energy efficiency, and performance under varying agricultural conditions. To address this gap, this study investigates the following research questions:

1. What are the key performance trade-offs (coverage, power consumption, data rate, cost) between LPWAN technologies (LoRaWAN, NB-IoT, Sigfox) and cellular networks (4G/5G) in different agricultural scenarios?
2. How can hybrid connectivity models combining LPWAN and 5G enhance reliability and cost-efficiency in smart agriculture?

To answer these questions, this paper systematically analyzes recent academic research (2020–2024) on IoT connectivity in smart agriculture, focusing on network performance optimization and hybrid architectures. By evaluating methodologies such as simulation, field trials, and analytical modeling, this study provides a comparative assessment of different connectivity technologies and offers evidence-based recommendations for their deployment in real-world agricultural applications.

The remainder of this paper is structured as follows: Section 2 outlines the research methodology, including evaluation criteria for IoT connectivity in smart agriculture. Section 3 reviews existing LPWAN and cellular network technologies, highlighting their advantages and limitations. Section 4 explores hybrid LPWAN-5G models and their impact on connectivity performance. Sections 5 discusses research gaps, future directions, and recommendations for optimizing IoT-based agricultural networks. Finally, Sections 6 concludes the paper.

## 2   Methodology

This study follows a systematic and structured approach to evaluate the performance trade-offs between Low-Power Wide-Area Networks (LPWAN) and cellular networks in smart agriculture, as well as to explore the potential of hybrid IoT connectiv-ity models. The research focuses on key connectivity parameters, including coverage, power consumption, data rate, latency, and cost, which are critical for IoT deployments in diverse agricultural environments. These thematic areas guide our analysis and ensure a comprehensive and structured evaluation of IoT connectivity solutions for smart farming.

To ensure an up-to-date and relevant review, we analyzed peer-reviewed literature published between 2020 and 2024, sourcing papers from IEEE Xplore, ACM Digital Library, ScienceDirect, and MDPI. Our search utilized targeted keywords such as "smart agriculture IoT," "LPWAN LoRaWAN NB-IoT Sigfox," "4G 5G agriculture IoT," and "LPWAN 5G integration". Preference was given to studies that:
- Explicitly compare network performance in agricultural applications.
- Propose novel connectivity frameworks or hybrid network architectures for smart farming.
- Include real-world agricultural deployments, particularly in tropical and remote farming environments when available.

In addition to research articles, survey papers and review studies were incorporated to provide broader contextual insights. The initial pool of studies was systematically filtered based on relevance to the research questions and the agricultural use cases covered. Each selected paper was examined to extract:
- Research objectives and scope.
- Methodological approach (e.g., simulation parameters, field experiment setups, analytical models).
- Key findings related to IoT connectivity performance.

Rather than merely summarizing existing studies, this paper emphasizes the methodologies used in previous research to provide an evidence-based assessment of connectivity technologies. The insights derived from this analysis serve as a foundation for evaluating the feasibility and efficiency of hybrid LPWAN–5G models in enhancing reliability and cost-effectiveness in smart agriculture.

## 3 LPWAN vs. Cellular Networks in Agriculture: Performance Trade-offs

Selecting the optimal IoT connectivity solution for agriculture requires a careful balance between coverage, power consumption, data rate (throughput), and cost. Existing studies have conducted both quantitative and qualitative comparisons of LPWAN technologies—such as LoRaWAN, NB-IoT, and Sigfox—against cellular networks (4G/5G) based on these critical parameters. This section synthesizes key findings from the literature to address the first research question, which examines the performance trade-offs between LPWAN and cellular networks in various agricultural scenarios. We analyze how each technology performs across different smart farming applications, ranging from large-scale field sensing to real-time video monitoring for precision agriculture. Additionally, we provide an overview of the methodologies employed in prior research, including simulation models, field experiments, and ana-lytical frameworks, to ensure a comprehensive and evidence-based assessment of connectivity solutions in smart agriculture.

### 3.1 Coverage and Range

LPWAN protocols are specifically designed for long-range communication, making them particularly suitable for smart agriculture applications in rural and remote areas where existing network infrastructure is limited [7, 8]. LoRaWAN and Sigfox can extend coverage over several kilometres, even in challenging terrains. For instance, a LoRaWAN deployment in the Andean highlands of Ecuador successfully achieved reliable connectivity across a 50-hectare farm, maintaining stable communication with sensor nodes located up to 875 meters away, despite obstructions from hilly terrain and dense vegetation [9]. Similarly, Sigfox has demonstrated long-range capabilities of up to 10 km in open environments, though its low data rate constrains its usability for applications requiring frequent or high-volume data transmissions [10].

The authors in [11] conducted an in-depth study on the integration of unmanned aerial vehicles (UAVs), LPWAN, and IoT technologies to enhance farm monitoring and smart agriculture applications. Their work specifically focuses on developing an IoT-based water quality and livestock monitoring systems, where a multi-channel LoRaWAN gateway is integrated into a UAV to relay sensor data to the cloud for further analysis. Through comprehensive measurements and simulations, they evaluated aerial LoRaWAN communication, identifying optimal configurations for long-range connectivity. Their findings demonstrate that UAV-assisted LoRaWAN significantly extends coverage (up to 10 km), ensures reliable communication even at high drone speeds, and optimizes aerial data collection by reducing flight times by a factor of four compared to traditional mission planning. This study provides valuable insights into overcoming the limitations of terrestrial LoRaWAN networks, offering a promising solution for large-scale smart farming deployments.

On the other hand, NB-IoT, as a cellular LPWAN technology, benefits from existing mobile network infrastructure. One study estimated that NB-IoT's transmission range extends several kilometers, utilizing coverage enhancement techniques and path-loss models (such as the Hata model) for rural environments. In real-world scenarios, NB-IoT cells can provide coverage comparable to LoRaWAN, especially when operating in low-frequency bands (e.g., 800 MHz) and leveraging coverage extension features [12].

Traditional 4G LTE networks provide reliable outdoor coverage, typically spanning a few kilometers per cell in rural macrocell deployments. However, connectivity in remote farms depends on the availability of operator infrastructure. In contrast, 5G networks exhibit varying range capabilities:

- **Sub-6 GHz 5G bands** offer an overall range similar to 4G, making them viable for wide-area farm connectivity.
- **5G mmWave**, while delivering high data rates, is impractical for agricultural use due to its limited range and susceptibility to signal obstruction.

In remote and sparsely connected rural regions, LPWAN technologies often serve as a cost-effective solution, particularly in areas where 4G/5G signals are weak or absent. Since LoRaWAN and Sigfox can be deployed as private networks, they are frequently used to fill coverage gaps in smart agriculture. LoRaWAN, in particular, has been identified as an effective alternative for connecting distant fields and greenhouses where deploying cellular infrastructure is impractical.

Overall, LPWAN technologies outperform traditional cellular networks in coverage per base station, making them ideal for large farms and remote plantations. In contrast, cellular networks (4G/5G) rely on existing tower deployments and may require additional infrastructure investments to extend connectivity into agricultural dead zones.

### 3.2 Power Consumption and Battery Life

One of the key advantages of LPWAN technologies for agricultural IoT is their ultra-low power consumption, which enables battery-powered sensors to operate for several years without maintenance. LoRaWAN and Sigfox devices are designed to transmit small data packets infrequently, allowing them to remain in deep sleep mode for extended periods, consuming only a few microamps when idle. Empirical studies confirm that LoRaWAN nodes can achieve multi-year battery life, particularly when spreading factors and transmission intervals are optimized [13]. Similarly, Sigfox devices, constrained to sending 12-byte messages up to 140 times per day, can also sustain years of operation on a single battery.

In contrast, NB-IoT devices exhibit higher power consumption due to network attachment requirements and higher transmission power, resulting in greater energy usage than LoRaWAN for similar tasks. A study by Ugwuanyi et al. [12] conducted a direct power consumption comparison between LoRaWAN and NB-IoT under controlled conditions, analyzing network joining energy, uplink transmission power, and overall battery life implications. Their findings revealed that NB-IoT modules consumed approximately 3 mAh to join the network, while LoRaWAN required only 1 mAh, making LoRaWAN three times more power-efficient in establishing connectivity. Additionally, a 44-byte uplink message in NB-IoT consumed 1.8 mAh, whereas LoRaWAN required only 100 µAh, resulting in an 18× higher energy cost per uplink transmission for NB-IoT. These results underscore LoRaWAN's superior battery longevity, particularly for periodic sensor readings where low-power operation is essential.

Traditional 4G and 5G networks, commonly used for high-bandwidth IoT applications, have significantly higher power demands, often requiring mains power or frequent recharging. Even LTE Cat-M1, an IoT-optimized variant of LTE, exhibits higher baseline power consumption than NB-IoT. As a result, battery-powered LPWAN sensors (LoRaWAN or NB-IoT) remain the preferred choice for large-scale smart agriculture deployments, particularly in remote areas with limited access to power infrastructure.

A KTH study (2020) [13] further estimated the battery lifetime of LoRaWAN vs. NB-IoT under periodic sensing conditions using a 3000 mAh battery. The findings suggested that:

- LoRaWAN nodes (SF7) could operate for approximately 32.7 months before requiring battery replacement.
- NB-IoT nodes, under similar conditions, lasted around 20 months, demonstrating lower energy efficiency despite its low-power optimizations.

Overall, LPWAN technologies provide significantly longer battery life than 4G/5G-based IoT devices, making them ideal for smart farming applications that require extended sensor deployments with minimal maintenance. Among LPWAN technologies, LoRaWAN and Sigfox typically outperform NB-IoT in terms of energy efficiency, though NB-IoT can still achieve multi-year battery life with proper duty cycling and optimized transmission strategies.

### 3.3 Data Rate and Capacity

A fundamental trade-off in LPWAN technologies is their limited data throughput, which is a consequence of their long-range and low-power operation. LoRaWAN data rates range from 0.3 kbps (at maximum spreading factor 12) to tens of kbps under optimal conditions. Sigfox is even more constrained, with typical uplink speeds of 100 to 600 bps and minimal downlink capacity, supporting only a few acknowledgments per day. As a result, Sigfox is primarily suited for ultra-low-bandwidth applications, such as simple sensor readings.

Among LPWAN options, NB-IoT provides the highest throughput, with peak downlink speeds of approximately 250 kbps and uplink rates between 20–60 kbps, as defined by 3GPP specifications. However, real-world throughput is often significantly lower (tens of kbps or less) due to narrowband operation and network overhead. A controlled evaluation found that NB-IoT achieved a maximum sustainable throughput of 264 bps with ~837 ms latency, exceeding LoRaWAN's throughput in the same test environment while maintaining robust performance under network congestion [14]. These findings suggest that NB-IoT can handle slightly higher data demands, such as frequent telemetry and small image transfers, more efficiently than LoRaWAN in a dense network scenario.

In contrast, 4G and 5G networks offer significantly higher data rates, enabling applications that require large data streams and low-latency communication. For instance, 5G networks in farm settings can facilitate high-definition video streaming from drone cameras, real-time machinery teleoperation, and AI-driven farm automation. One study demonstrated that 5G-enabled industrial livestock farming leveraged centralized data aggregation to connect thermostats, feeding systems, and herd monitoring sensors, optimizing resource management and livestock welfare. Additionally, 5G enhances drone-based agricultural monitoring, supporting fast, high-quality video transmission for precision farming applications [15]. These capabilities surpass traditional 4G/Wi-Fi-based systems in speed, scalability, and reliability, making autonomous agricultural vehicles, predictive analytics, and large-scale environmental monitoring more feasible. However, such high-bandwidth connectivity is unnecessary for simpler applications, such as soil moisture sensing or basic weather monitoring, where LPWAN remains the more efficient option.

Network capacity also varies across these technologies. LoRaWAN operates in unlicensed spectrum, making it susceptible to duty cycle limitations that can restrict scalability in areas with thousands of deployed nodes. In contrast, NB-IoT and cellular networks handle significantly more devices per cell, benefiting from scheduled access mechanisms. A study by Levchenko et al. [10] analyzed packet loss under network congestion for Sigfox, LoRaWAN, and NB-Fi (a newer LPWAN alternative). Their findings showed that:

- Sigfox performed best with infrequent, small messages, given its strict duty cycle constraints.
- LoRaWAN was most reliable for transmitting larger data chunks, though duty-cycle regulations still posed limitations.
- NB-Fi (and by extension NB-IoT) excelled in scenarios requiring frequent acknowledgments, improving overall quality of service (QoS).

These insights suggest that for large-scale agricultural deployments with thousands of sensors (e.g., extensive greenhouse complexes or nationwide environmental monitoring), NB-IoT's integration into cellular networks offers greater scalability. However, this comes at the cost of higher per-node energy consumption, making LoRaWAN or Sigfox preferable for ultra-low-power applications in remote areas.

### 3.4 Cost and Deployment Considerations

Cost is a critical factor influencing farmers' adoption of IoT solutions, particularly in resource-constrained agricultural settings. LPWAN technologies typically offer lower upfront and operational costs, especially when private deployments are feasible. For instance, a LoRaWAN network can be established by deploying a small number of gateway devices (each costing a few hundred euro) connected to an internet backhaul. Since LoRaWAN operates in the license-free ISM spectrum, there are no recurring spectrum fees, making it an attractive option for developing regions and research initiatives. For example, agricultural communities in Bangladesh have increasingly adopted LoRaWAN due to its cost-effective long-range coverage and low energy consumption, eliminating reliance on expensive telecom infrastructure [16].

Sigfox, on the other hand, requires a subscription to a service operator, though its per-device costs remain low (typically a few euro per year). Unlike LoRaWAN, Sigfox users do not need to maintain gateway hardware, simplifying deployment but introducing dependency on third-party networks.

NB-IoT and cellular-based IoT solutions depend on mobile network operators, meaning their cost-effectiveness varies by location. In urbanized regions, NB-IoT can leverage existing 4G coverage with minimal additional infrastructure costs. However, in rural areas with poor mobile network availability, expanding cellular networks is expensive, requiring new base stations. NB-IoT devices also require SIM cards and data plans, though specialized IoT packages are often more affordable than standard mobile subscriptions.

5G infrastructure costs are significantly higher, particularly for large-scale agricultural deployments. Deploying 5G base stations across extensive farms or plantations is prohibitively expensive unless integrated into nationwide rural broadband initia-tives. Consequently, the return on investment (ROI) for 5G in agriculture depends on the scale and complexity of applications—it is justifiable for advanced use cases such as autonomous farm vehicles, precision agriculture, and large-scale sensor grids but less practical for basic farm monitoring tasks.

A comparative analysis in Bangladesh by Haque et al. [16] highlighted that while 5G provides the most advanced capabilities, its high initial infrastructure costs and potential operating expenses make simpler technologies (LoRaWAN or even 4G) more cost-effective for basic farm monitoring applications. In contrast, 4G networks offer a balance between affordability and performance, benefiting from mature infrastructure and widespread device availability. Many smart farming implementations continue to rely on 3G/4G-based telemetry, with farmers using mobile internet to receive sensor data, primarily due to extensive cellular coverage and relatively low operating costs.

In summary, choosing the right IoT connectivity solution depends on deployment conditions. LPWAN solutions (LoRaWAN, Sigfox) offer cost advantages where cellular infrastructure is limited, while cellular IoT (NB-IoT, 4G, and 5G) may be preferable in regions with strong mobile network coverage. For large-scale sensor networks, subscription-based models may scale poorly, making private LPWAN deployments more economical in the long run. Table 1 summarizes the cost factors for NB-IoT, LoRAWAN, and Sigfox deployment.

**Table 1.** Cost factors for NB-IoT, LoRAWAN, and Sigfox deployment.

| Cost Factor for Deployment | NB-IoT | Sigfox | LoRaWAN (Own local network) | LoRaWAN (Subscription) |
|---|---|---|---|---|
| Devices incl radio module | € | € | € | € |
| SIM card | € | - | - | - |
| Subscription fee | € | € | - | € |
| Own network infrastructure | - | - | € | - |
| Network ops & maintenance | - | - | € | - |
| Application (servers) | € | € | € | € |

### 3.5 Summary of Trade-offs

Table 2 provides a comparative overview of the key characteristics of LPWAN and cellular IoT technologies, synthesizing data from multiple sources to frame the subsequent performance evaluation. This comparison highlights how different connectivity solutions balance range, data rate, power consumption, and cost, serving as a foundation for the comparative analysis presented in this study.

Table 2 demonstrates that LPWAN technologies (LoRaWAN, Sigfox, NB-IoT) prioritize extended coverage and energy efficiency at the expense of data rates, making them well-suited for sensor-based monitoring applications that require small, infrequent data transmissions over large agricultural areas. In contrast, cellular networks (4G/5G) offer significantly higher data rates and lower latency, which are essential for high-bandwidth applications such as drone-based imaging, video surveillance, and real-time machinery control. However, these high-performance capabilities come with greater power consumption and higher deployment costs, making them less suitable for battery-operated, large-scale sensor networks.

Table 2. Key characteristics of LPWAN vs. cellular technologies for IoT in agriculture.

| Technology | Frequency Band | Typical Range | Data Rate | Power/Battery Life | Network Cost Model |
|---|---|---|---|---|---|
| LoRaWAN (LPWAN) | Unlicensed ISM (≈868/915 MHz) | Up to 10-15 km rural; 2–5 km urban | ~0.3-50 kbps (adaptive SF) | Ultra-low power; battery life can exceed 10 years with optimized transmission intervals. | Private network; low infrastructure cost; no subscription fee. |
| Sigfox (LPWAN) | Unlicensed ISM (≈868/915 MHz) | Up to ~10 km rural; 3-5 km urban | ~100-600 bps (tiny payloads) | Ultra-low power; battery life up to 10 years (with infrequent messages). | Subscription-based; requires Sigfox network operator. |
| NB-IoT (LPWAN) | Licensed LTE band (e.g., 800 MHz) | ~1-10 km (cellular coverage; +20 dB link budget for deep coverage) | up to 250 kbps downlink and 20–60 kbps uplink. | Low power (PSM/eDRX enable multi-year, but higher than LoRa). | Operated by mobile carriers; SIM card and data plan required. |
| 4G LTE | Licensed bands (700 MHz-2.6 GHz) | ~1-5 km per cell (macro-cell) | up to 100 Mbps downlink and 50 Mbps uplink. | Higher power consumption than LPWAN; battery life ranges from days to months depending on usage. | Subscription-based; SIM card and data plan required. |
| 5G NR (sub-6 GHz) | Licensed sub-6 GHz (<6 GHz) | ~0.5-3 km per cell (sub-6 GHz macro-cell); | Up to 10 Gbps (eMBB), up to 250 kbps (mMTC), optimized for low latency (URLLC) | Higher power for eMBB devices (smartphones); IoT devices (RedCap/NB-IoT) optimized for low power. | Subscription-based; requires SIM card and data plan. |

The optimal connectivity choice varies depending on the specific agricultural application. For instance:
- A soil moisture sensor network in paddy fields would benefit from LoRaWAN or Sigfox, as they provide low-cost, long-range connectivity with multi-year battery life.
- A fleet of precision spraying drones, which requires real-time video streaming and command-and-control functionality, would be better served by 5G due to its high throughput and low latency.

Researchers commonly employ a combination of simulation models and field trials to evaluate these trade-offs. For example, Soy et al. [12] developed two prototype vehicle tracking units (VTUs)—one using LoRaWAN and the other using NB-IoT—and analyzed their coverage performance across urban, suburban, and rural environments. Their study derived closed-form analytical expressions for maximum transmission range using the Hata path loss model, and they further validated their results using the XIRIO Online radio-frequency planning tool. Such empirical methodologies provide strong validation for the theoretical performance advantages of each technology, reinforcing the importance of application-specific connectivity selection.

The consensus in recent literature is that no single network technology is universally "best" for all smart farming applications. Instead, the choice of connectivity must be carefully tailored to the specific trade-offs between coverage, throughput, power efficiency, and cost. This realization has driven increasing interest in hybrid IoT connectivity solutions, which combine multiple technologies to leverage their complementary strengths—such as integrating LPWAN for long-range sensing with 5G for high-speed, low-latency data processing.

## 4      Hybrid LPWAN–5G Connectivity Models for Smart Agriculture

The complementary nature of LPWAN and cellular networks has led to the emergence of hybrid connectivity models designed to enhance reliability, coverage, and cost-efficiency in agricultural IoT deployments. These hybrid architectures directly address Research Question 2: How can combining LPWAN technologies (e.g., LoRaWAN, NB-IoT) with 5G improve smart agriculture networks?

The core concept behind hybrid LPWAN–5G integration is to utilize LPWAN for energy-efficient, long-range sensing, while leveraging 5G (or 4G) as a high-bandwidth backhaul for real-time data processing and high-speed communication. This approach ensures that low-power, battery-operated sensors can collect and transmit essential agricultural data over long distances, while 5G enables rapid aggregation, edge computing, and AI-driven analytics for precision farming, drone-based monitoring, and autonomous machinery control. The following subsections provide a detailed analysis of this integration from different aspects.

### 4.1 Architectural Approaches

A common hybrid LPWAN–5G architecture follows a two-tier network model, as illustrated in Figure 1, where resource-constrained IoT sensors connect via LPWAN to a local gateway, which then communicates with cloud platforms via 4G/5G [17]. The gateway acts as a bridge, linking the low-power local network to the high-speed cellular infrastructure. This model is already widely implemented; for instance, a LoRaWAN gateway on a farm can use a 4G SIM card to aggregate and transmit sensor data to a cloud server, thereby combining LoRa's long-range sensor coverage with cellular internet access

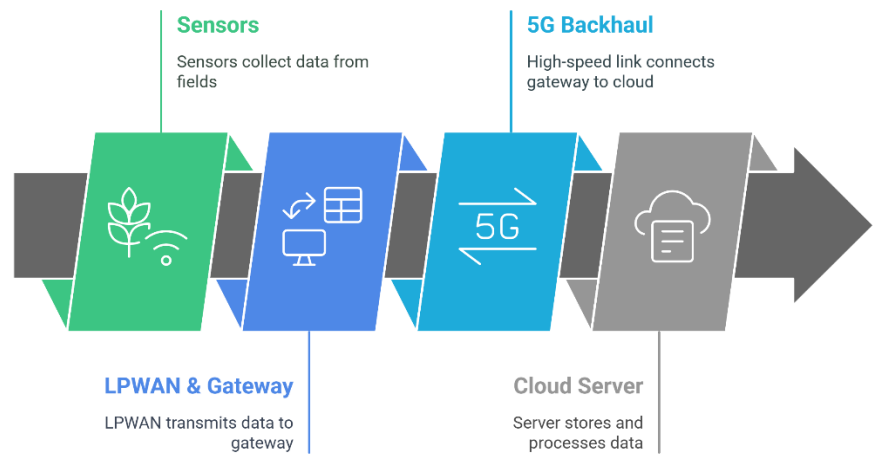

**Fig 1.** Hybrid LPWAN-5G Architecture with 5G as Backhaul.

Figure 2 illustrates an alternative hybrid LPWAN–5G architecture, where both LPWAN and 5G operate in parallel for different data collection purposes in smart agriculture. Unlike the backhaul-based integration in Figure 1, this model leverages the strengths of each network independently, enabling a multi-tiered sensing and communication approach. In this setup:

- Simple, low-power IoT sensors (e.g., soil moisture, temperature, or air quality sensors) use LPWAN (LoRaWAN, NB-IoT, or Sigfox) to transmit low-data-rate environmental information to a LPWAN gateway, which then forwards aggregated data to the cloud server or 5G base stations.
- Advanced sensors and high-bandwidth devices (e.g., drones for real-time crop imaging and autonomous farming machinery) rely on 5G connectivity via a 5G base station to enable low-latency, high-throughput data transfer directly to the cloud.

This parallel architecture optimizes network efficiency, allowing LPWAN to handle routine, energy-efficient monitoring while 5G supports mission-critical, data-intensive applications. The division of tasks enhances reliability, scalability, and cost-effectiveness, as only devices requiring high-speed, real-time communication consume 5G bandwidth, while low-power sensors operate over cost-efficient LPWAN links.

This model is particularly useful for large-scale smart farming deployments, where heterogeneous sensor networks must support varied data requirements—from long-range, battery-powered monitoring to real-time analytics-driven automation.

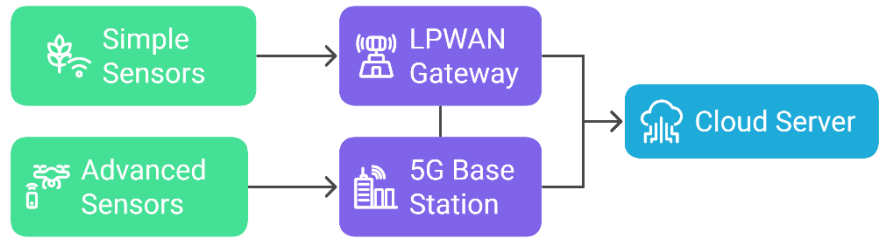

**Fig 2.** Hybrid LPWAN-5G with Parallel Data Collection.

Ogbodo et al. [6] conducted a comprehensive survey on hybrid 5G–LPWAN architectures, emphasizing the direct integration of 5G with LoRaWAN and NB-IoT to enhance ubiquitous IoT connectivity. They propose leveraging 5G's software-defined networking (SDN) and network function virtualization (NFV) capabilities to create virtual network slices optimized for LPWAN traffic. This network slicing approach enables 5G networks to efficiently handle large-scale, low-power IoT deployments, while ensuring Quality of Service (QoS) for diverse applications.

The authors advocate for a unified 5G–LPWAN IoT architecture, integrating emerging technologies such as SDN, NFV, and mobile edge computing (MEC) to support both urban and remote-area IoT applications. In this model, 5G networks serve as the backhaul for LPWAN gateways, facilitating seamless communication between IoT devices and the core network. NB-IoT, which operates in licensed cellular bands, is highlighted as a key enabler for low-power, wide-area IoT applications within the 5G ecosystem.

To extend connectivity to underserved remote areas, Ogbodo et al. propose integrating LPWAN with non-terrestrial networks (NTNs), such as Low-Earth Orbit (LEO) satellites. They highlight the 5G NB-IoT NTN standard (introduced in 3GPP Release 17) as a promising solution for IoT deployments in regions beyond the reach of terrestrial networks. This approach ensures continuous connectivity for isolated locations, such as remote farms or maritime environments, allowing them to benefit from advanced IoT applications despite geographical constraints.

The study in [18] deployed a hybrid LPWAN wireless connectivity solution for lake-based water quality monitoring, demonstrating that a combination of Smart Utility Network (SUN), LoRaWAN, and WiFi—assisted by a low-altitude platform (balloon)—effectively overcame non-line-of-sight (NLOS) transmission constraints. The system successfully relayed sensor data over a ground distance of 2.4 km by utilizing multiple SUN and LoRa nodes. These findings highlight the effectiveness of hybrid network architectures in addressing connectivity challenges in remote environments, a strategy that can be similarly applied to smart agriculture, particularly in remote and hard-to-reach regions.

Recent advancements in hybrid IoT connectivity have introduced novel architectures that extend the applicability of LPWAN–5G models to smart farming. One such approach [19] integrates Aerial Access Networks (AAN), Federated Learning (FL), and hybrid LoRa Point-to-Point (P2P)/LoRaWAN technologies to enhance data communication, processing, and predictive modeling in large-scale, remote, and hard-to-reach agricultural environments. Conventional smart monitoring systems often struggle with long-distance data transmission, high operational costs, and limited adaptability to dynamic environmental conditions, especially in areas with poor telecommunications infrastructure. The proposed hybrid system addresses these limitations by utilizing AAN for extended coverage, FL for decentralized and privacy-preserving data analytics, and hybrid LoRa P2P/LoRaWAN for efficient and scalable connectivity. Preliminary simulation results confirm the feasibility and effectiveness of this architecture in environmental monitoring, highlighting its potential application in smart farming for precision agriculture, autonomous irrigation control, and large-scale crop health assessment. This approach demonstrates a paradigm shift in agricultural IoT by enabling real-time, adaptive, and energy-efficient monitoring solutions, making it a promising direction for future research and deployment in precision farming.

### 4.2 Reliability and Redundancy

Hybrid LPWAN–5G models significantly enhance reliability by providing multiple communication paths, ensuring continuous operation even in the presence of network failures or environmental interference. For example, if an on-farm LoRaWAN network experiences interference or a gateway failure, a 5G backup link can take over for critical applications, such as emergency irrigation shutoff systems or high-value asset tracking. Although this fallback mechanism incurs higher power consumption, it ensures that essential functions remain operational. Conversely, if the cellular network suffers downtime or weak coverage in certain areas of a farm, the LPWAN network can continue operating locally, storing and forwarding data once connectivity is restored.

To further enhance network resilience, some implementations employ multi-homing strategies, where IoT nodes or gateways dynamically switch between LPWAN and cellular networks based on real-time network conditions and power availability. This adaptability is crucial for agricultural operations that rely on continuous data collection and alerts, such as greenhouse climate control and automated irrigation management.

Edge computing at network gateways further improves reliability and efficiency in hybrid architectures. Bhattacharya & De [20] introduced AgriEdge, an edge-intelligent framework that integrates 5G, NB-IoT, and drones for precision farming applications. In this model:
- Drones collect high-resolution agricultural data, such as crop health imagery.

414- Edge processors (located at farm servers or base stations) perform real-time analysis, reducing data transmission overhead.
- NB-IoT (an LPWAN mode under 5G) transmits summarized insights to the cloud, ensuring efficient bandwidth utilization and low-latency decision-making.

By processing data locally, AgriEdge reduces bandwidth consumption, lowers network latency, and optimally allocates 5G for high-data applications (e.g., drone imagery processing) while utilizing LPWAN for low-data telemetry (e.g., soil moisture monitoring). This type of hybrid edge architecture enables real-time agricultural decision-making, such as identifying crop stress from drone images via 5G while soil sensors concurrently report moisture levels via LPWAN.

The methodology behind such frameworks typically involves developing prototypes or simulations that incorporate both network types, followed by performance evaluations based on latency, data delivery success rates, and energy efficiency. Bhattacharya & De validated AgriEdge through simulations, demonstrating its feasibility and effectiveness for real-time precision farming. Their findings underscore the value of combining 5G's high-bandwidth capabilities with LPWAN's energy-efficient connectivity, reinforcing hybrid architectures as a robust solution for next-generation smart agriculture.

### 4.3 Cost Efficiency

From a cost-efficiency perspective, hybrid LPWAN–5G models can optimize resource allocation by reserving expensive 5G bandwidth for applications that truly require high data rates and low latency, while offloading bulk, low-priority traffic to cost-effective LPWAN networks. For instance, a large-scale deployment of soil moisture sensors could use LoRaWAN to communicate with a local gateway, incurring no recurring telecommunications fees, while high-value assets such as autonomous tractors or machine vision systems leverage 5G for real-time data transmission. This selective use of 5G ensures that deployments remain economically scalable while maintaining optimal performance for diverse agricultural IoT applications [6].

A recent study by Jradi et al. [21] proposes a seamless integration of LoRaWAN into 5G networks, demonstrating how this hybrid approach significantly reduces operational costs by utilizing LoRaWAN's cost-efficient infrastructure, which operates in unlicensed spectrum bands. Their proposed system employs LoRaWAN nodes for local sensing and data aggregation, while gateways act as intermediaries, transmitting aggregated data via cellular backhaul (LTE or 5G) to cloud platforms. By minimizing the number of required cellular subscriptions, this model substantially lowers operational and deployment costs, making it particularly beneficial for large-scale IoT applications in expansive agricultural regions.

In regions, where agricultural lands are vast and budgets are constrained, such hybrid architectures offer a practical solution. A 2020 mangrove forest monitoring project in Malaysia [22] provides a real-world example of such integration. Jebril et al. developed a hybrid LPWAN architecture, using LoRa-based sensor nodes for image capture, an edge gateway for preprocessing, and an NB-IoT uplink for cloud trans-mission. Their design overcame LoRa's bandwidth limitations by splitting image data and forwarding it via NB-IoT whenever available, achieving a balance between energy efficiency and data delivery for remote camera sensors. This study demonstrates how hybrid LPWAN setups (LoRa + NB-IoT) can efficiently handle high-bandwidth tasks, such as image transmission, while maintaining energy efficiency in remote monitoring applications.

Methodologies for evaluating cost efficiency in these hybrid models often involve comparative analyses or network simulations, such as:
- Calculating data plan costs for pure 5G deployments versus hybrid LPWAN–5G networks.
- Simulating network usage scenarios, including how frequently dual-mode devices switch to cellular links versus relying on free LPWAN communication.

These studies consistently highlight that offloading even a portion of IoT traffic to unlicensed LPWAN networks can result in substantial cost savings, particularly in large-scale deployments where device counts are high.

### 4.4    Challenges in Integration:

While LPWAN–5G integration offers significant advantages, it also presents several technical and operational challenges. Interfacing disparate technologies requires careful network design, as gateways or base stations must translate between different communication protocols, potentially introducing latency if not efficiently managed. Additionally, security remains a key concern, as ensuring end-to-end data protection across LPWAN and 5G networks—each with distinct security standards—necessitates a unified and robust security framework.

Mobility is another critical challenge. LPWAN technologies such as LoRaWAN are designed for static or slow-moving devices, lacking seamless handover capabilities between gateways. In contrast, cellular networks are optimized for mobility, enabling continuous connectivity for moving devices. A hybrid tracking system could address this by using LoRaWAN for static sensors (e.g., weather stations, soil monitors) while employing 5G for mobile applications such as cattle tracking or autonomous agricultural vehicles, ensuring uninterrupted data transmission.

Ogbodo et al. [6] highlight the need for ubiquitous connectivity, proposing that future 5G networks could natively support LPWAN protocols, enabling seamless transitions between network types. However, real-world implementations of hybrid networks in agriculture remain limited, with most deployments still in testbed or pilot stages. The consensus in recent studies is that hybrid models significantly enhance reliability, coverage, and cost efficiency by leveraging the strengths of each technology. However, they require advanced network management solutions to handle protocol translation, security, and resource allocation effectively.

As 5G networks continue expanding into rural areas the adoption of hybrid LPWAN–5G architectures in smart farming is expected to grow. Farms will increasingly rely on local LPWANs for energy-efficient sensor monitoring, while utilizing 5G links for high-data applications, ultimately achieving comprehensive connectivity and improved agricultural productivity.

## 5 Methodologies in Current Research

The surveyed literature employs a variety of methodologies to evaluate and optimize IoT connectivity for smart agriculture. Common approaches include:

### 5.1 Simulation and Theoretical Modeling

Many studies employ simulation and theoretical modeling as key methodologies to evaluate and optimize IoT connectivity for smart agriculture. Network simulators and analytical models are frequently used to predict performance metrics such as coverage, capacity, and latency. For instance, path-loss models like Okumura-Hata have been applied to analytically derive coverage ranges for LoRaWAN and NB-IoT across different terrains [12].

Other studies focus on large-scale network behavior simulations. For example, Levchenko et al. [10] conducted simulations involving thousands of IoT nodes to assess the performance of LPWAN technologies, including NB-Fi, Sigfox, and LoRaWAN, under varying traffic loads. Their analysis focused on key network performance metrics, such as packet loss rate (PLR), packet error rate (PER), and average delay, providing valuable insights into the trade-offs between coverage, capacity, and network congestion.

Simulations are particularly useful for extrapolating performance metrics beyond what is feasible to test in real-world conditions. However, a notable limitation of this approach is its potential oversimplification of real-world agricultural environments. Many simulations assume uniform terrain, static interference conditions, and evenly distributed sensor nodes, whereas actual farm deployments are subject to dynamic interference, terrain irregularities, and non-uniform sensor placement. These real-world complexities can significantly impact network performance, suggesting that simulation results should be complemented with field experiments for a more comprehensive evaluation of IoT connectivity in smart agriculture.

### 5.2 Experimental Testbeds and Field Trials

Several studies have employed prototype networks and field testbeds to gather empirical data on IoT connectivity performance in agricultural environments. Ugwuanyi et al. [22] developed a dual testbed to compare LoRaWAN and NB-IoT under controlled conditions. Their LoRaWAN setup used a Pysense node as both a gateway and a sensor node, operating on the 868 MHz frequency band with a coding rate of 4/5, 125 kHz bandwidth, and a spreading factor (SF) of 7. For NB-IoT, the testbed incorporated a LimeSDR (LMS7002M) to generate LTE core MME and eNB signals, with a Fipy and Pysense expansion board as the NB-IoT node, configured for in-band operation on Band 28 (780.500 MHz DL and 725.500 MHz UL). These setups allowed for comprehensive performance measurement in a controlled environment, providing insights into network behavior and efficiency.

Another field study [9] deployed a LoRa network in agricultural settings in the Andean region of Ecuador, focusing on mountainous areas above 2,910 meters above sea level. This research evaluated Packet Delivery Ratio (PDR), Received Signal

Strength Indicator (RSSI), and Signal-to-Noise Ratio (SNR) under varying environmental conditions, including terrain slopes, vegetation density, and weather variations (sun, rain, wind). The results showed that in areas with 15% slopes, surrounded by eucalyptus trees, and without line-of-sight (NLoS):

- RSSI values ranged from -111 dBm to -122 dBm, indicating moderate signal attenuation.
- SNR values varied between -2.55 dB and -15.14 dB, reflecting interference and terrain-induced degradation.
- Uplink communications achieved a PDR no lower than 76%, while downlink PDR exceeded 92%, even in challenging environmental conditions.

These field trials demonstrate the significant impact of terrain, vegetation, and weather on LoRaWAN performance, offering valuable insights for network planning in complex agricultural landscapes.

Experimental testbeds provide real-world validation and help uncover practical deployment challenges that theoretical models may overlook. For example, antenna placement, power consumption under different conditions, and weather-induced signal degradation can all significantly influence network performance. Common methodologies in field trials involve logging network performance metrics over extended periods (days or weeks) and varying operational parameters (e.g., data transmission interval, spreading factor) to assess their impact.

However, a key challenge in field trials is reproducibility and scalability. Many experiments are conducted in a single location with a limited number of devices, making it difficult to generalize results across diverse agricultural settings. Future research should aim to expand field trial coverage, incorporating multi-location studies and larger-scale deployments to better capture the variability of real-world farming environments.

### 5.3 Comparative Surveys and Reviews

Some studies adopt a survey-based approach, systematically reviewing prior research to synthesize key findings and trends in IoT connectivity for smart agriculture. Pagano et al. [23] and Ogbodo et al. [6] conducted comprehensive surveys, compiling technical specifications and reported performance metrics of various IoT technologies, with a focus on LoRaWAN, 5G, and LPWAN applications. Pagano et al. examined LoRaWAN's role in smart agriculture, identifying key challenges such as downlink latency, energy management, and device heterogeneity. Their review also explored the potential of machine learning and edge computing to enhance the scalability and efficiency of LoRa-based agricultural IoT systems.

Conversely, Ogbodo et al. provided a broader review of IoT technologies, emphasizing the integration of LPWAN with 5G for smart cities and agriculture. They highlighted the lack of a unified architecture for 5G–LPWAN integration and the need for advanced security mechanisms, including endogenous security frameworks to address vulnerabilities in hybrid networks. Additionally, they advocated for the use of Low Earth Orbit (LEO) satellites to enable ubiquitous IoT connectivity in remote and hard-to-reach regions.

The authors in [5] provided an extensive literature review on wireless IoT technologies, focusing on the key challenges in IoT deployments, including power consumption, quality of service, localization, and wireless channel propagation modeling. Their study particularly emphasizes the importance of accurate channel modeling and characterization, which is crucial for optimizing IoT network performance and enabling efficient deployment in diverse environments. By summarizing recent advancements in wireless channel modeling, they offer a generalized framework for commonly used propagation models, providing valuable insights for network planners. Their work serves as a foundational guideline for future research on wireless IoT deployment and channel modeling strategies, ensuring robust and efficient network design in emerging IoT applications.

Methodologically, these survey studies rely on content analysis of existing literature, drawing generalized conclusions about the current state and future directions of IoT technologies in their respective domains. These reviews offer valuable insights and identify emerging trends and play a crucial role in identifying research gaps, guiding future investigations, and shaping the development of more robust IoT architectures for smart agriculture.

### 5.4 Use-case Demonstrations

Several studies focus on IoT applications in smart agriculture, implementing connectivity solutions tailored to specific use cases such as precision irrigation and livestock tracking. These studies combine network engineering with agricultural performance metrics, evaluating both technical parameters (e.g., network reliability, latency) and real-world impact (e.g., water efficiency, crop yield improvement).

For example, Zhang et al. [24] developed an IoT-based precision irrigation system using LoRaWAN to optimize water management for cabbage crops. The system integrated soil moisture sensors (SWC and SWP), solenoid valves, and a LoRaWAN gateway for real-time monitoring and automated irrigation control via the AllThingsTalk platform. Four irrigation strategies were tested, but the experiment was cut short due to cold weather, limiting data collection on water efficiency and crop yield. Despite this, the study demonstrated LoRaWAN's potential for precision irrigation, while identifying challenges such as 4.3% data packet loss over 300 meters and high power consumption in continuous sensor operation.

Such use cases highlight how IoT connectivity solutions can enhance agricultural efficiency, while also revealing practical deployment challenges that must be addressed for large-scale implementation.

## 6 Research Gaps and Future Directions

### 6.1 Limited Multi-Technology Comparisons

While some studies compare specific IoT technologies (e.g., LoRa vs. NB-IoT, Sigfox vs. NB-Fi), there is a lack of comprehensive, side-by-side evaluations covering all major LPWAN and cellular options under identical conditions. No single field trial has concurrently tested LoRaWAN, Sigfox, NB-IoT, LTE-M, and 5G on the same farm, making it difficult to fully quantify trade-offs. As a result, researchers must often cross-compare findings from different papers, which may be based on varying assumptions and methodologies.

For example, Soy et al. [12] conducted a theoretical and simulation-based analysis of LoRa and NB-IoT, using the Hata path-loss model to estimate coverage ranges in urban, suburban, and rural environments. However, their study lacks real-world field trials and does not consider other LPWAN and cellular technologies like Sigfox, LTE-M, or 5G, limiting its practical applicability. Similarly, Ugwuanyi et al. [22] performed real-world testbed evaluations of LoRaWAN and NB-IoT, comparing latency, throughput, and power consumption. While this study offers empirical insights, it excludes Sigfox, LTE-M, and 5G, leaving gaps in understanding how these technologies perform in a unified setup.

The key gap in current research is the absence of a large-scale, real-world comparative study that evaluates all major LPWAN and cellular technologies under the same environmental and operational conditions. Such a study would provide valuable insights into coverage, energy efficiency, latency, and scalability, enabling better-informed decisions for IoT deployments in agriculture and other sectors. Additionally, the limited number of real-world field trials constrains the ability to validate theoretical models, emphasizing the need for empirical research that directly benchmarks different connectivity solutions in agricultural settings.

### 6.2   Lack of Long-Term Evaluations

Agricultural IoT deployments must function reliably over extended periods, enduring seasonal variations, extreme weather conditions (e.g., rain, storms), and multi-year operational lifecycles. However, most existing studies and field experiments are short-term, typically lasting only days or weeks, limiting insights into long-term network durability and performance stability.

A notable gap in current research is the absence of longitudinal studies that evaluate IoT connectivity over months or years. Such studies would provide valuable insights into real-world challenges, including battery degradation in sensors, network maintenance requirements, and connectivity stability across multiple planting and harvesting cycles. Without long-term data, it is difficult to assess how well IoT solutions sustain performance under changing agricultural conditions.

Long-term empirical evaluations are crucial for gaining stakeholder confidence, particularly among farmers and agribusinesses considering large-scale IoT adoption. Future research should prioritize extended trials to validate the real-world reliability, scalability, and cost-effectiveness of IoT-enabled smart farming solutions.

### 6.3   Integration and Interoperability Testing

While hybrid IoT network architectures have been widely proposed (as discussed in previous sections), real-world implementations remain in their early stages. A significant gap exists in interoperability testing methodologies, particularly in ensuring seamless handovers between LPWAN and cellular networks, such as LoRaWAN transitioning to 4G/5G, or aggregating data from heterogeneous networks into a unified platform.

Interoperability standards—potentially facilitated by IoT middleware or multi-protocol gateways—are still emerging, and there is limited research on fully integrated LPWAN–5G deployments. Most existing hybrid setups rely on manual configurations, where LoRaWAN networks feed data into a 4G gateway, rather than implementing automated, dynamic switching mechanisms.

Future research should focus on developing interoperability frameworks and prototypes, ensuring seamless communication between different IoT protocols. Achieving this requires combining hardware and software stacks from multiple domains, a complex but necessary step toward scalable, robust, and intelligent agricultural IoT ecosystems.

### 6.4 Scalability and Network Congestion in Large-Scale Deployments

While LPWAN technologies like LoRaWAN, NB-IoT, and Sigfox are designed for massive IoT connectivity, real-world large-scale agricultural deployments may face scalability challenges due to network congestion, interference, and duty-cycle limitations. LoRaWAN, for example, operates in unlicensed spectrum bands with duty-cycle restrictions, meaning that as the number of connected devices increases, packet collisions and transmission delays may degrade performance. Similarly, NB-IoT and LTE-M, while leveraging licensed spectrum, may experience capacity constraints in rural areas with limited base station coverage.

Future studies should investigate scalability optimization techniques for large-scale smart farming IoT networks. This includes:
- Adaptive MAC protocols to manage network congestion dynamically.
- Multi-tier network architectures, where local LPWAN networks aggregate data before transmission to cellular or satellite backhaul.
- AI-driven resource allocation mechanisms to prioritize mission-critical agricultural data traffic while minimizing packet collisions and energy waste.

Developing efficient congestion control strategies will be critical for enabling massive IoT connectivity in large agricultural regions.

### 6.5 Energy Harvesting and Self-Sustaining IoT Nodes

IoT deployments in agriculture often rely on battery-powered sensors, which require periodic maintenance and replacement, increasing operational costs and logistical challenges—especially in remote or large-scale farms. While low-power techniques (e.g., duty cycling, adaptive transmission rates) can extend battery life, they do not eliminate the need for periodic battery changes. The lack of self-sustaining power solutions is a major bottleneck for long-term IoT deployments in agriculture.

A promising solution is to explore energy harvesting techniques to develop self-sustaining IoT nodes for smart farming. Future research can focus on:
- Solar, wind, and RF energy harvesting integrated with LPWAN devices.
- Hybrid energy systems, where IoT nodes dynamically switch between battery power and harvested energy based on environmental conditions.

- Ultra-low-power hardware designs combined with energy-efficient network protocols to minimize energy consumption further.

Developing self-sustaining, maintenance-free IoT nodes would greatly enhance the feasibility of large-scale smart agriculture deployments.

### 6.6 Edge AI for Real-Time Agricultural Decision-Making

Most IoT-based smart agriculture solutions rely on cloud computing for data processing, introducing latency and bandwidth constraints, especially in rural or remote areas with limited connectivity. Transmitting raw sensor data to the cloud can result in high data costs, network congestion, and increased response times, which may not be ideal for real-time agricultural applications, such as automated irrigation, disease detection, or livestock monitoring.

Future studies should focus on integrating AI-driven edge computing into agricultural IoT systems to enable:

- On-device data processing for real-time anomaly detection (e.g., plant disease, soil moisture stress).
- Federated learning-based models to allow distributed AI training on IoT nodes without requiring constant cloud connectivity.
- Edge-cloud hybrid architectures, where only high-priority insights are sent to the cloud, reducing network load and latency.

Implementing Edge AI in IoT-enabled smart farming would allow faster decision-making, reduce reliance on cloud infrastructure, and improve overall system efficiency.

The current body of research on IoT connectivity for smart agriculture presents significant advancements but also reveals key limitations that must be addressed for scalable, efficient, and reliable deployments. The lack of comprehensive multi-technology comparisons, long-term evaluations, and real-world interoperability testing limits our understanding of network performance under diverse agricultural conditions. Additionally, the underrepresentation of emerging LPWAN technologies, alongside a predominant focus on network metrics rather than agricultural impact, highlights the need for more application-driven research. Future studies should focus on scalability solutions for large deployments, self-sustaining energy harvesting IoT nodes, and AI-powered edge computing to enable real-time decision-making and long-term operational viability. Addressing these gaps will be crucial for advancing smart agriculture technologies and ensuring practical, cost-effective, and region-specific IoT implementations.

## 7 Conclusion

This study reviewed IoT connectivity solutions for smart agriculture, comparing LPWAN (LoRaWAN, NB-IoT, Sigfox) and cellular networks (4G/5G). LPWAN excels in coverage and energy efficiency, making it ideal for low-power, long-range monitoring, while 4G/5G offers high throughput and low latency for data-intensive applications like drone imaging and autonomous farming. No single technology meets all agricultural needs, but hybrid LPWAN–5G models provide a scalable and cost-effective solution, leveraging LPWAN for routine data and cellular for high-priority transmissions.

Empirical studies confirm that hybrid architectures enhance network reliability and efficiency, particularly when integrated with edge computing and network slicing. However, research gaps remain in long-term evaluations, multi-technology comparisons, and region-specific deployments. Future work should assess network performance over extended periods and explore localized IoT implementations in diverse agricultural environments.

This review provides a technical roadmap for optimizing IoT in agriculture, guiding stakeholders in balancing coverage, power efficiency, data capacity, and cost to accelerate sustainable and intelligent farming practices.